\documentclass[12pt]{article}
\usepackage{amsmath}
\usepackage{amssymb}
\begin{document}
\begin{titlepage}
\begin{flushright}
hep-th/0612267\\
TIT/HEP-564\\
YITP-06-65\\
December, 2006\\
\end{flushright}
\vspace{0.5cm}
\begin{center}
{\Large \bf 
Deformation of $\mathcal{N}=4$ Super
Yang-Mills Theory in Graviphoton Background
}
\lineskip .75em
\vskip2.5cm
{\large Katsushi Ito${}^{1}$, Yoshishige Kobayashi${}^{1}$ and Shin 
Sasaki${}^{2}$ }
\vskip 2.5em
${}^{1}$ {\large\it Department of Physics\\
Tokyo Institute of Technology\\
Tokyo, 152-8551, Japan}  \vskip 1.5em
${}^{2}$ {\large\it Yukawa Institute for Theoretical Physics \\
Kyoto University \\
Kyoto 606--8502, Japan} \vskip 4.5em
\end{center}
\begin{abstract}
We study deformation of $\mathcal{N}=4$ super Yang-Mills
theory  from type IIB superstrings  with D3-branes in 
the constant R-R background.
We compute disk amplitudes with
one graviphoton vertex operator
and investigate the zero-slope limit of the amplitudes.
We obtain the effective action deformed by the graviphoton background,
which contains the one defined in non(anti)commutative ${\cal N}=1$ 
superspace as special case.
The bosonic part of the Lagrangian gives the Chern-Simons term
coupled with the R-R potential.
We study the vacuum configuration of the deformed Lagrangian and
find the fuzzy sphere configuration for scalar fields.

\end{abstract}
\end{titlepage}

\baselineskip=0.7cm
\section{Introduction}
Ramond-Ramond (R-R) background in type II superstring theories 
plays an important role in studying effective theories on the 
D-branes.
In particular constant graviphoton background induces
non(anti)commutativity in world-volume superspace \cite{OoVa, DeGrNi, BeSe}.
$\mathcal{N}=1$ super Yang-Mills 
theory on non(anti)commutative superspace ($\mathcal{N}=1/2$ superspace)
\cite{Se} 
is obtained from type IIB superstrings with constant graviphoton
background compactified on a Calabi-Yau threefold.
The deformed action was constructed explicitly in \cite{Billo-1/2} from
open string amplitudes in type IIB superstring theory compactified 
on an orbifold ${\bf C}^3/{\bf Z}_2\times{\bf Z}_2$ in graviphoton
background. 

Non(anti)commutative $\mathcal{N}=2$ harmonic superspace
provides various types of deformed $\mathcal{N}=2$ supersymmetric
gauge theories \cite{IvLeZu,ArItOh2,FeIvLeSoZu,ArIt3,Castro}.
In a previous paper \cite{ItSa}, 
two of the present authors studied the deformation
of $\mathcal{N}=2$ supersymmetric Yang-Mills theory from the open string
disk amplitudes with one graviphoton vertex operator in type IIB superstring
theory.
The deformed $\mathcal{N}=2$ super Yang-Mills theory is  realized as
the low-energy effective theory on the D3-branes in the type 
IIB superstrings compactified on ${\bf C}^2/{\bf Z}_2$ with constant 
graviphoton background.
The constant R-R backgrounds ${\cal F}^{\alpha\beta ij}$ are
 classified into four types of
deformations (S,S), (S,A), (A,S) and (A,A)-types, in which (S,S) and
(A,A)-types deformations are related to the deformation of
$\mathcal{N}=2$ superspace.
By choosing the (S,S)-type graviphoton background and the appropriate scaling
condition, it was shown that
the effective Lagrangian on the D3-branes becomes the
deformed one in non(anti)commutative ${\cal N}=2$ harmonic superspace
at the lowest order in deformation parameters.

It is an interesting problem 
to extend this deformation to $\mathcal{N}=4$ case.
Since superspace formalism keeping $\mathcal{N}=4$ supersymmetry manifestly
 is not yet known,
superstring approach provides a systematic
method for understanding general non(anti)commutative deformation of
$\mathcal{N}=4$ theory.
The couplings between R-R fields and the world-volume massless fields on
the D$p$-branes have been studied in \cite{My, GaMy}, which are
written in the form of the Chern-Simons action.
In recent papers \cite{Im2}, by restricting 
the constant five-form background to the deformation parameter of 
$\mathcal{N}=1/2$ superspace, it is shown that the bosonic part of 
the Chern-Simons action reduces to the deformed interaction terms of the
$\mathcal{N}=4$ super Yang-Mills theory in $\mathcal{N}=1/2$ superspace.
The interaction terms including fermions are constructed by
the supersymmetric completion \cite{Im2} using remaining supersymmetries
but disagree with the deformed action based on non(anti)commutative
superspace.

In this paper we will study the deformation of $\mathcal{N}=4$ 
supersymmetric Yang-Mills
theory from open string amplitudes in the constant R-R background.
We put the D3-branes in type IIB superstrings in flat ten-dimensional
(Euclidean) space-time.
We will compute open superstring disk amplitudes with one graviphoton vertex 
operator and determine the low-energy effective Lagrangian at the first
order in the deformation parameter $C$.
We will find the effective action in a special graviphoton background
is consistent with the deformed action defined on non(anti)commutative
$\mathcal{N}=1$ superspace at the first order in $C$.

The action contains new scalar potential terms deformed by $C$.
This term corresponds to the Myers term \cite{My} which gives rise to a
dielectric configuration for scalar fields.
In the ${\cal N}=1/2$ superspace formalism, this type of configuration
was found in \cite{Im2}. 
In the present paper, we will explore the fuzzy two-sphere configuration
for the scalar fields in the presence of constant $U(1)$ gauge fields,
based on the (S,S)-type deformed action.

This paper is organized as follows:
in section 2, we review D3-brane realization of $\mathcal{N}=4$
super Yang-Mill theory in type IIB superstring theory and classify the 
R-R background.
In section 3, we 
calculate the disk amplitudes with one closed string graviphoton vertex
operator and the effective action on the D3-branes.
In section 4, we study the vacuum configuration of the deformed 
$\mathcal{N}=4$ theory and find the fuzzy two-sphere configuration for
scalar fields.

\section{D3-brane realization of  $\mathcal{N}=4$ super Yang-Mills theory}

In this section we review the D3-brane realization of ${\cal N}=4$ super
Yang-Mills theory \cite{BFPFLL}.

\subsection{Type IIB superstrings}
We begin with explaining type IIB superstrings in flat space-time.
We will use the NSR formalism.
Let $X^{m}(z,\bar{z})$, $\psi^m(z)$ and $\tilde{\psi}^m(\bar{z})$ 
($m=1,\cdots,10$)
be free bosons and
fermions with world-sheet coordinates $(z,\bar{z})$.
Their operator product
expansions (OPEs) are given by
$X^m(z)X^n(w)\sim -\delta^{mn}\ln(z-w)$ and
$\psi^m(z)\psi^n(w)\sim \delta^{mn}/(z-w)$.
Here the space-time signature is Euclidean.
Fermionic ghost system $(b,c)$ with conformal weight ($2,-1$) 
and bosonic ghost system $(\beta,\gamma)$ with weight ($3/2,-1/2$) are
also introduced. 
The world-sheet fermions $\psi^m(z)$ are bosonized in terms of free
bosons $\phi^a(z)$ $(a=1,\cdots,5)$ by
\begin{eqnarray}
 f^{\pm e_a}(z)&\equiv&{1\over\sqrt{2}}(\psi^{2a-1}\mp i \psi^{2a})
=: e^{\pm \phi^a}(z): c_{e^a}.
\end{eqnarray}
Here $\phi^a(z)$ satisfy the OPE $\phi^a(z)\phi^b(w)\sim
\delta^{ab}\ln(z-w)$ and the vectors $e_a$ are orthonormal basis 
 in the $SO(10)$ weight lattice space and
$c_{e^a}$ is a cocycle factor \cite{KoLeLeSaWa}.
The bosonic ghost is also bosonized \cite{FMS}:
$
\beta=\partial\xi e^{-\phi}$,
$
\gamma= e^{\phi}\eta
$
with OPE $\phi(z)\phi(w)\sim -\ln(z-w)$. 
The R-sector is constructed from spin fields
$S^{\lambda}(z)=e^{\lambda\phi}(z)c_{\lambda}$, where $\phi = \phi^a 
e_a$ and $\lambda
={1\over2}(\pm e_1\pm e_2\pm e_3\pm e_4\pm e_5)$.
 $\lambda$ belongs to the spinor representation of $SO(10)$.
$c_{\lambda}$ is a cocycle factor.
In type IIB theory, after the GSO projection, 
we have spinor fields which have 
odd number of minus signs in $\lambda$, 
for both left and right movers.

We now introduce parallel $N$ D3-branes in the $(x^1,\cdots,x^3)$-directions.
Since the D3-branes breaks the ten-dimensional Lorentz symmetry 
$SO(10)$ to $SO(4)\times SO(6)$,
the spin field $S^{\lambda}(z)$ is decomposed as
$(S_\alpha S_A, S^{\dot{\alpha}}S^A)$, where $S_\alpha$ and
$S^{\dot{\alpha}}$ ($\alpha,\dot{\alpha}=1,2$) are four-dimensional Weyl 
spinor and
$S_{A}$ and $S^{A}$ ($A=1,2,3,4$) are six-dimensional Weyl spinor.
$S^A$ ($S_A$) belongs to the (anti-)fundamental representation of
$SU(4)$.
The Dirac matrices for four dimensional part
are
$\sigma_\mu=(i\tau^1,i\tau^2,i\tau^3,1)$ and
$\bar{\sigma}_\mu=(-i\tau^1,-i\tau^2,-i\tau^3,1)$,
where $\tau^i$ ($i=1,2,3$) are the Pauli matrices.
The Lorentz generators are defined by
$
\sigma^{\mu \nu} = \frac{1}{4}
( \sigma^{\mu} \bar{\sigma}^{\nu} - \sigma^{\nu} \bar{\sigma}^{\mu})
$
and
$
\bar{\sigma}_{\mu\nu}={1\over4}(\bar{\sigma}_\mu\sigma_\mu-\bar{\sigma}_\nu
\sigma_\mu)
$.

The gamma matrices for six-dimensional part are
given by
\begin{equation}
\Sigma^a=
\left( \eta^{3}, - i\bar{\eta}^{3}, \eta^{2}, - i\bar{\eta}^{2},
\eta^{1}, i\bar{\eta}^{1} \right),
\quad \bar{\Sigma}^a=
(-\eta^3, -i\bar{\eta}^3, -\eta^2,-i\bar{\eta}^2, -\eta^1, i\bar{\eta}^1), 
\label{eq:sixgamma}
\end{equation}
where $a=1,\cdots,6$. $\eta^a_{\mu\nu}$ and $\bar{\eta}^a_{\mu\nu}$ are
{}'t Hooft symbols, which are defined by
$\sigma_{\mu\nu}={i\over2}\eta^a_{\mu\nu} \tau^a$ and
$\bar{\sigma}_{\mu\nu}={i\over2}\bar{\eta}^a_{\mu\nu} \tau^a$.
The matrices (\ref{eq:sixgamma}) satisfy the algebra
\begin{equation}
(\Sigma^a
 )^{AB}(\bar{\Sigma}^a)_{BC}+(\Sigma^b)^{AB}(\bar{\Sigma}^a)_{BC}
=2\delta^{ab}\delta^A_C .
\end{equation}

The massless spectrum of open strings contain gauge fields $A_\mu$, six
scalars $\varphi^a$ in the NS sector and gauginos $\Lambda^{\alpha A}$
and $\bar{\Lambda}_{\dot{\alpha} A}$ in the R sector.
The vertex operators for gauge and scalar fields in the $(-1)$ picture 
are given by
\begin{eqnarray}
& & V_A^{(-1)} (y;p) = (2 \pi \alpha')^{\frac{1}{2}} \frac{A_{\mu} 
 (p)}{\sqrt{2}} \psi^{\mu} (y) e^{- \phi (y)} e^{i \sqrt{2 \pi \alpha'} p
 \cdot X (y)}, \nonumber \\
& &V_{\varphi}^{(-1)} (y;p) = (2 \pi \alpha')^{\frac{1}{2}} \frac{\varphi_a 
 (p)}{\sqrt{2}} \psi^a (y) e^{- \phi (y)} e^{i \sqrt{2 \pi \alpha'} p
 \cdot X (y)}, \nonumber \\ \label{vo-1}
\end{eqnarray}
while in the $0$ picture, they are given by
\begin{eqnarray}
& &V_A^{(0)} (y;p) = 2 i (2 \pi \alpha')^{\frac{1}{2}} A_{\mu} (p) 
 \left( \partial X^{\mu} (y) + i (2 \pi \alpha')^{\frac{1}{2}} p \cdot 
  \psi \psi^{\mu} (y)  \right) e^{i \sqrt{2 \pi \alpha'} p \cdot X (y)},
\nonumber\\
& &V_{\varphi}^{(0)} (y;p) = 2 i (2 \pi \alpha')^{\frac{1}{2}} \varphi_a (p) 
 \left( \partial X^a (y) + i (2 \pi \alpha')^{\frac{1}{2}} p \cdot 
  \psi \psi^a (y)  \right) e^{i \sqrt{2 \pi \alpha'} p \cdot X (y)}. \label{vo-2}
\end{eqnarray}
The gaugino vertex operators in the $(-1/2)$ picture are
\begin{eqnarray}
& &V_{\Lambda}^{(-1/2)} (y;p) = \frac{}{} (2 \pi \alpha')^{\frac{3}{4}} 
 \Lambda^{\alpha A} (p) S_{\alpha} (y) S_A (y) e^{- \frac{1}{2} \phi 
 (y)} e^{i \sqrt{2 \pi \alpha' } p \cdot X (y)}, \nonumber \\
& &V_{\overline{\Lambda}}^{(-1/2)} (y;p) = \frac{}{}( 2 \pi \alpha' 
 )^{\frac{3}{4}} \overline{\Lambda}_{\dot{\alpha} A} (p) S^{\dot{\alpha}} (y) 
 S^A (y) e^{-\frac{1}{2} \phi (y)} e^{i \sqrt{2 \pi \alpha' } p 
 \cdot X (y)}. \label{vo-3}
\end{eqnarray}
We adapt dimensionless four-momentum $\sqrt{2 \pi \alpha'} p$ to ensure 
that the momentum representation of a field have same dimension of 
space-time field.

\subsection{R-R vertex operator}
The vertex operators for massless states
in the R-R sector of type IIB superstrings are
constructed from the tensor product of spin fields $(S^\alpha S^A,
S_{\dot{\alpha}}S_A)$ and $(\tilde{S}^{\beta}\tilde{S}^B, \tilde{S}_{\dot{\beta}}\tilde{S}_{B})$.
We will study the effect of constant R-R background, which is described by 
the closed string vertex operator
\begin{eqnarray}
V_{\mathcal{F}}^{(-1/2,-1/2)} (z, \bar{z}) = 
(2 \pi \alpha')^{\frac{3}{2}} \mathcal{F}^{\alpha \beta A B}
S_{\alpha} S_A e^{- \frac{1}{2} \phi }(z) 
\tilde{S}_{\beta} \tilde{S}_B e^{- \frac{1}{2} 
\tilde{\phi} }( \bar{z}). \label{R-R_vertex_operator}
\end{eqnarray} 
We note that general massless closed string states in the R-R sector
contain the field strength of types
$\mathcal{F}^{\alpha}{}_{\dot{\alpha}} {}^{A}{}_B$,
$\mathcal{F}_{\dot{\alpha}}{}^{\alpha} {}_A{}^{B}$,
and
$\mathcal{F}_{\dot{\alpha}\dot{\beta}AB}$.
Since the vertex operator (\ref{R-R_vertex_operator}) provides a
generalization of the 
deformations of $\mathcal{N}=2$  non(anti)commutative superspace
\cite{ItSa}, we will consider this type of deformation in the present work.

As in the $\mathcal{N}=2$ case \cite{ItSa}, 
the field strength is decomposed as
\begin{eqnarray}
\mathcal{F}^{\alpha \beta AB} = \mathcal{F}^{(\alpha \beta) (AB)}
+ \mathcal{F}^{(\alpha \beta) [AB]} + \mathcal{F}^{[\alpha \beta] (AB)}
+ \mathcal{F}^{[\alpha \beta] [AB]}.
\end{eqnarray}
Here the parenthesis $(AB)$ represents symmetrization of indices $A$ and
$B$. 
$[AB]$ represents anti-symmetrization. 
We call these backgrounds  
(S,S), (S,A), (A,S), (A,A)-type deformations respectively.
We now examine the correspondence between the field strength 
$\mathcal{F}^{\alpha \beta A B}$ and 
the $p$-form R-R field strengths in type IIB superstrings.

For four-dimensional sector, the tensor $f^{\alpha\beta}$ can be
decomposed into the singlet and the self-dual tensor parts:
\begin{equation}
f^{\alpha\beta}
= \epsilon^{\alpha \beta} f
+ (\sigma^{\mu \nu} )^{\alpha \beta} f_{\mu \nu}.
\label{eq:fourform}
\end{equation}
The first term corresponds to antisymmetric part and the second to the
symmetric part.

For six-dimensional sector, the spinor indices are labeled by the
fundamental representation $\underline{\bf 4}$ of $SU(4)$.
The tensor product $\underline{\bf 4}\otimes \underline{\bf 4}$
can be decomposed into $\underline{\bf 6}\oplus \underline{\bf 10}$,
which corresponds to the vector or the self-dual 3-form representation, 
respectively.
In fact, the tensor $g^{AB}$ is expressed as
\begin{equation}
g^{AB}=
(\Sigma^{a})^{AB} g_{a} +
(\Sigma^{abc})^{AB} g_{abc}.
\label{eq:sixform}
\end{equation}
Here we define a matrix which is totally
antisymmetric with respect to the space indices of $abc$,
\begin{equation}
(\Sigma^{abc})^{AB} \equiv ( \Sigma^{[a}\bar{\Sigma}^{b}\Sigma^{c]})^{AB}.
\end{equation}
The first term in (\ref{eq:sixform}) corresponds to the antisymmetric
part and the second to the symmetric part.
The matrix $\Sigma^{abc}$ is self-dual:
\begin{equation}
(\Sigma^{abc})^{AB} = \frac{i}{3!} \epsilon^{abcdef} (\Sigma_{def})^{AB},
\end{equation}
and consequently the three-form $g_{abc}$ also
satisfies the self-dual condition,
\begin{equation}
g_{abc} = \frac{i}{3!} \epsilon_{abcdef} g^{def}.
\end{equation}

{}From the decompositions (\ref{eq:fourform}) and (\ref{eq:sixform}), we
find that the 
$\mathcal{F}^{\alpha \beta A B}$ can be decomposed into
\begin{eqnarray}
\mathcal{F}^{\alpha \beta A B } &= &
\left( \epsilon^{\alpha \beta} f
+ (\sigma^{\mu \nu} )^{\alpha \beta} f_{\mu \nu}
\right)
\times
\left( (\Sigma^{a})^{AB} \; g_{a} +
(\Sigma^{abc})^{AB} \; g_{abc} \right) \nonumber \\
&= &
f g_a \epsilon^{\alpha\beta}(\Sigma^{a})^{AB}
+f g_{abc} \epsilon^{\alpha\beta} (\Sigma^{abc})^{AB}
+ f_{\mu \nu} g_{a} (\sigma^{\mu \nu} )^{\alpha \beta} (\Sigma^{a})^{AB}
\nonumber\\
&&+ f_{\mu \nu} g_{abc} (\sigma^{\mu \nu} )^{\alpha \beta}
(\Sigma^{abc})^{AB},
\label{eq:decomp}
\end{eqnarray}
which corresponds to 
\begin{equation}
\mathcal{F}^{\alpha \beta AB} \thicksim (\text{R-R 1-form})\oplus(\text{R-R 3-form})\oplus
(\text{R-R 3-form})\oplus (\text{R-R 5-form}).
\end{equation}
The decomposition (\ref{eq:decomp}) shows that
the (A,A) deformation corresponds to the R-R 1-form,
the (A,S) and (S,A) deformations to the R-R 3-forms, and
the (S,S) deformation to the R-R self-dual 5-form.
In fact, if we identify the self-dual five-form field strength $F^{mnpqr}$ as
\begin{equation}
F^{\mu \nu a b c} = f_{\mu \nu} g_{abc},
\end{equation}
then it satisfies the self-dual condition in the 10-dimensional space,
\begin{equation}
F^{\mu \nu a b c} =
\frac{i}{2!3!} \epsilon^{\mu \nu a b c \rho \sigma d e f} 
F_{\rho \sigma d e f}.
\end{equation}
 
We note that the similar decomposition holds in the case of
the deformation of $\mathcal{N}=2$ super Yang-Mills theory \cite{ItSa},
which is constructed from the type IIB superstrings compactified on
${\bf C}\times {\bf C}^2/{\bf Z}^2$.

\subsection{Disk amplitudes and auxiliary field method}
The action of $\mathcal{N}=4$ supersymmetric Yang-Mills theory is 
obtained by evaluating correlation functions of the vertex operators 
given in equations (\ref{vo-1}), (\ref{vo-2}), (\ref{vo-3}).
Let us consider disk amplitudes with boundary attached on the D3-brane 
world volume.
The disk is realized as the upper half of complex plane.
The boundary condition of the 
spin field \cite{BFPFLL} is
\begin{equation}
\left. S_{\alpha} S_A (z)= \tilde{S}_{\alpha} \tilde{S}_A (\bar{z})
\right|_{z=\bar{z}} .
\end{equation}
The disk amplitudes can be calculated by replacing
$\tilde{S}_{\alpha} \tilde{S}_A (\bar{z})$ by
$S_{\alpha} S_A (\bar{z})$ in the correlator.
The $n+2 n_{\cal F}$-point disk amplitude for $n$ vertex operators
$V^{(q_i)}_{X_i}(y_i)$ and $n_{\cal F}$ R-R vertex operators
$V^{(-{1\over2},-{1\over2})}_{\cal F}(z_j,\bar{z}_j)$
is given by
\begin{equation}
 \langle \! \langle V^{(q_1)}_{X_1}\cdots V^{(-{1\over2},-{1\over2})}_{\cal
  F}
\cdots \rangle \! \rangle
=C_{D_2}\int {\prod_{i=1}^{n}dy_i \prod_{j=1}^{n_{\cal F}}
dz_jd\bar{z}_j
\over dV_{CKG}}
\langle V^{(q_1)}_{X_1}(y_1)\cdots V^{(-{1\over2},-{1\over2})}_{\cal F}
(z_1,\bar{z_1})\cdots \rangle.
\label{eq:disk1}
\end{equation}
Here $C_{D_2}$ is the disk normalization factor
\cite{VMLRM}:
\begin{equation}
 C_{D_2}={1\over 2\pi^2(\alpha')^2 }{1\over k g_{\mathrm{YM}}^2}
\end{equation}
and $g_{\mathrm{YM}}$ is the gauge coupling constant. $k$ is a
normalization constant of $U(N)$ generators
$T^a$: ${\rm Tr}(T^a T^b)=k\delta^{ab}$.
$dV_{CKG}$ is an $SL(2,{\bf R})$-invariant volume factor to fix
three positions $x_1$, $x_2$ and $x_3$ among $y_i$, $z_j$,and
$\bar{z}_j$'s:
\begin{equation}
 d V_{CKG}={d x_1 d x_2 d x_3\over
(x_1-x_2) (x_2-x_3) (x_3-x_1)}.
\end{equation}

The open string amplitudes in the zero slope limit 
shows that the effective action on the D3-branes is that of
$\mathcal{N}=4$ super Yang-Mills theory:
\begin{eqnarray}
\mathcal{L}_{\mathrm{SYM}}^{\mathcal{N} = 4} &=& \frac{1}{k} \frac{1}{g^2_{\mathrm{YM}}} \mathrm{Tr} \left[ 
- \frac{1}{4} F^{\mu\nu} \left(F_{\mu \nu} + \tilde{F}_{\mu \nu} \right) 
- i \Lambda^{\alpha A} (\sigma^{\mu})_{\alpha \dot{\beta}} D_{\mu} 
\overline{\Lambda}^{\dot{\beta}}_{\ A}
    - \frac{1}{2} \left(D_{\mu} \varphi_a \right)^2 \right. \nonumber \\
& & \left. + \frac{1}{2} \left( \Sigma^a \right)^{AB} \overline{\Lambda}_{\dot{\alpha}A}
[\varphi_a, \overline{\Lambda}^{\dot{\alpha}}_{\ B} ] + \frac{1}{2} \left( 
\overline{\Sigma}^a \right)_{AB} \Lambda^{\alpha A} [\varphi_a, 
\Lambda_{\alpha}^B ] + \frac{1}{4} [\varphi_a, \varphi_b]^2   \right],
\label{N=4SYM}
\end{eqnarray}
where
\begin{eqnarray}
 F_{\mu\nu}&=& \partial_\mu A_\nu-\partial_\nu A_\mu+i [A_\mu, A_\nu], 
  \nonumber \\
D_{\mu}\varphi_a &=&\partial_\mu \varphi_a + i[A_\mu,\varphi_a],
\end{eqnarray}
and $A_\mu=A_{\mu}^a T^a$ etc. 
$\tilde{F}_{\mu\nu}$ is the dual of $F_{\mu\nu}$. 

We use the auxiliary field method
\cite{Billo-1/2, ItSa} to simplify  the string amplitudes including
contact terms.
We introduce the auxiliary fields $H_{\mu\nu}$,
$H_{\mu a}$ and $H_{ab}$ and rewrite  the action (\ref{N=4SYM}) 
into the form
\begin{eqnarray}
\mathcal{L}_{\mathrm{SYM}} &=& -  \frac{1}{g^2_{\mathrm{YM}}} \frac{1}{k} \mathrm{Tr} \left[ \frac{1}{2} (\partial_{\mu} A_{\nu} - \partial_{\nu} 
A_{\mu}) \partial^{\mu} A^{\nu} +  i \partial_{\mu} A_{\nu} [A^{\mu}, A^{\nu}]
+ \frac{1}{2} H_c H^c + \frac{1}{2} H_c \eta^c_{\mu \nu} [A^{\mu}, A^{\nu}] \right] 
\nonumber \\
& &  -  \frac{1}{g^2_{\mathrm{YM}}} \frac{1}{k} \mathrm{Tr} 
\left[ \frac{1}{2} H_{ab} H_{ab} + \frac{1}{\sqrt{2}} H_{ab} [\varphi_a, \varphi_b] \right] \nonumber \\
& &  -  \frac{1}{g^2_{\mathrm{YM}}} \frac{1}{k} \mathrm{Tr} 
\left[ \frac{1}{2} \partial_{\mu} \varphi_a \partial^{\mu} \varphi^a + i \partial_{\mu} \varphi_a [A^{\mu}, \varphi_a] 
+ \frac{1}{2} H_{\mu a} {H}^{a \mu} + 
 H_{\mu a} [A^{\mu}, \varphi_a]  \right] \nonumber \\
& &  - \frac{1}{g^2_{\mathrm{YM}}} \frac{1}{k} \mathrm{Tr} \left[  
 i \Lambda^A \sigma^{\mu} D_{\mu} \bar{\Lambda}_A - \frac{1}{2} \left( \Sigma^a \right)^{AB} \overline{\Lambda}_{\dot{\alpha}A}
[\varphi_a, \overline{\Lambda}^{\dot{\alpha}}_{\ B} ] - \frac{1}{2} \left( 
\overline{\Sigma}^a \right)_{AB} \Lambda^{\alpha A} [\varphi_a, 
\Lambda_{\alpha}^B ]  \right]. \label{auxiliary_lagrangian}
\end{eqnarray}
All quartic interactions in (\ref{N=4SYM}) are replaced by cubic ones. 
The vertex operators for auxiliary fields are given by
\begin{eqnarray}
 V_{H_{AA}}^{(0)}(y) &=& \frac{1}{2} (2\pi \alpha') H_{\mu\nu}(p)\psi^\nu\psi^\mu
e^{i\sqrt{2\pi\alpha'}p\cdot X}(y), \nonumber \\
V_{H_{A \varphi}}^{(0)} (y;p) &=& 2 (2 \pi \alpha') 
H_{\mu a} (p)  \psi^{\mu} \psi^{a} (y) e^{i \sqrt{2 \pi \alpha'} p \cdot X (y)}, \nonumber \\
V_{H_{\varphi \varphi}}^{(0)} (y;p) &=& - \frac{1}{\sqrt{2}} (2 \pi \alpha') 
H_{ab} (p) \psi^{a} \psi^{b} (y) e^{i \sqrt{2 \pi \alpha'} p \cdot X (y)}.
\end{eqnarray}

\section{Disk amplitudes in the constant graviphoton background}
In this section
 we calculate disk amplitudes including  one graviphoton vertex 
operator in the zero-slope limit and 
study the deformed $\mathcal{N}=4$ super Yang-Mills  action 
at the order $\mathcal{O} (\mathcal{F})$.

As in the $\mathcal{N}=1$ \cite{Billo-1/2} and 
$\mathcal{N}=2$ \cite{Billo-N=2,ItSa} cases,
the deformed action depends on the scaling condition
for the graviphoton field strength.
In this paper we fix the zero-slope scaling of R-R field strength as
\begin{eqnarray}
(2 \pi \alpha')^{\frac{3}{2}} \mathcal{F}^{\alpha \beta A B} \equiv 
 C^{\alpha \beta AB} = \mathrm{fixed}.
\label{eq:scaling}
\end{eqnarray}
In this scaling, the parameter $C^{\alpha\beta A B}$ has mass dimension
$-1$, which is the same dimension as the deformation parameters in
non(anti)commutative superspace.
We will also focus on 
the (S,S)-type background $\mathcal{F}^{(\alpha\beta) (AB)}$,
which corresponds to the self-dual R-R 5-form background and is expected 
to give a generalization of  non-singlet deformation of $\mathcal{N}=2$
superspace \cite{ItSa}.

When the R-R vertex operator (\ref{R-R_vertex_operator}) is inserted in
the disk, the charge conservation for internal spin fields 
restricts possible insertions of the open string vertex operators.
In fact, the operators of types $\overline{\Lambda} 
\overline{\Lambda}$, $\Lambda \overline{\Lambda} \varphi$ and 
$\varphi \varphi \varphi$ cancel the internal charge of the R-R vertex 
operator. 
In the zero slope limit with the scaling condition (\ref{eq:scaling}),
we find that the following amplitudes become nonzero in the $(S,S)$-type
background:
\begin{eqnarray}
& & \langle \! \langle V^{(0)}_A V^{(-1/2)}_{\overline{\Lambda}} 
V^{(-1/2)}_{\overline{\Lambda}} 
  V^{(-1/2,-1/2)}_{\mathcal{F}} \rangle \! \rangle + 
\langle \! \langle V^{(0)}_{H_{AA}} V^{(-1/2)}_{\overline{\Lambda}} 
V^{(-1/2)}_{\overline{\Lambda}} 
  V^{(-1/2,-1/2)}_{\mathcal{F}} \rangle \! \rangle,  \label{amp1}\\
& & \langle \! \langle V^{(-1/2)}_{\Lambda} V^{(-1/2)}_{\overline{\Lambda}} 
V^{(0)}_{\varphi} V^{(-1/2,-1/2)}_{\mathcal{F}} \rangle \! \rangle + 
\langle \! \langle V^{(-1/2)}_{\Lambda} V^{(-1/2)}_{\overline{\Lambda}} 
V^{(0)}_{H_{\varphi \varphi}} V^{(-1/2,-1/2)}_{\mathcal{F}} \rangle \! 
\rangle,  \label{amp2}\\
& & \langle \! \langle V^{(0)}_{\varphi} V^{(0)}_{\varphi} V^{(-1)}_{\varphi} 
 V^{(-1/2,-1/2)}_{\mathcal{F}} \rangle \! \rangle + \langle \! \langle V^{(0)}_{H_{A \varphi}} 
V^{(0)}_{\varphi} V^{(-1)}_{\varphi} V^{(-1/2,-1/2)}_{\mathcal{F}} 
\rangle \! \rangle \nonumber \\
& & \qquad  + \langle \! \langle V^{(0)}_{H_{A \varphi}} 
V^{(0)}_{H_{A \varphi}} V^{(-1)}_{\varphi} 
 V^{(-1/2,-1/2)}_{\mathcal{F}} \rangle \! \rangle, \label{amp3}\\
& & \langle \! \langle V^{(0)}_{A} V^{(0)}_{H_{\varphi 
\varphi}} V^{(-1)}_{\varphi} V^{(-1/2,-1/2)}_{\mathcal{F}} \rangle \! 
\rangle + \langle \! \langle V^{(0)}_{H_{A\varphi}} V^{(0)}_{H_{\varphi
 \varphi 
}} V^{(-1)}_{\varphi} V^{(-1/2,-1/2)}_{\mathcal{F}} \rangle \! \rangle,
 \label{amp4}\\
& & \langle \! \langle V^{(0)}_{H_{\varphi \varphi}} V^{(-1/2)}_{\Lambda}
V^{(-1/2)}_{\Lambda} V^{(-1/2,-1/2)}_{\mathcal{F}} \rangle \! \rangle.
\label{amp5}
\end{eqnarray}
As in the  $\mathcal{N} = 1$ \cite{Billo-1/2} and $\mathcal{N} = 
2$ \cite{ItSa} cases, gauge invariance in the effective action is ensured 
by  the fact that the derivative $\partial_\mu A_\nu$ or
 $\partial_\mu\varphi$
appears together 
with auxiliary fields $H_{\mu\nu}$ and $H_{A\varphi}$, respectively.
The derivative terms turn out to be
covariant derivatives after integrating out auxiliary fields.
Appropriate weight factors of the amplitudes 
must be taken into account to keep the gauge invariance of the results.
We now compute the amplitudes (\ref{amp1})--(\ref{amp4}) explicitly.
\\
\\
\underline{\textbullet $\langle \! \langle V_A V_{\overline{\Lambda}} 
 V_{\overline{\Lambda}} V_{\mathcal{F}} \rangle \! \rangle$ + $\langle \! 
 \langle V_{H_{AA}} V_{\overline{\Lambda}} V_{\overline{\Lambda}} V_{\mathcal{F}} 
 \rangle \! \rangle$ }
\\
\\
The first term of the amplitudes (\ref{amp1}) is given by
\begin{eqnarray}
& & \langle \! \langle V^{(0)}_A (p_1) V^{(-1/2)}_{\overline{\Lambda}} (p_2) 
 V^{(-1/2)}_{\overline{\Lambda}} (p_3) V^{(-1/2,-1/2)}_{\mathcal{F}} 
\rangle \! \rangle 
 \nonumber \\
& & = \frac{1}{2 \pi^2 \alpha'^2} \frac{1}{k g^2_{\mathrm{YM}}}
 (2 i) (2 \pi \alpha')^{3} \mathrm{Tr} \left[ A_{\mu} (p_1)
\overline{\Lambda}_{\dot{\alpha} C} (p_2) \overline{\Lambda}_{\dot{\beta} 
D} (p_3) \right] \mathcal{F}^{(\alpha \beta) (AB)} \nonumber \\
& & \qquad \times \int \! \frac{\prod_j d y_j d z d \bar{z}}{d V_{\mathrm{CKG}}} \
\langle e^{-\frac{1}{2} \phi (y_1)} e^{-\frac{1}{2} \phi (y_2)}
 e^{-\frac{1}{2} \phi (z)} e^{-\frac{1}{2} \phi (\bar{z})} \rangle 
  \langle S^C (y_1) S^D (y_2) S_A (z) S_B (\bar{z}) \rangle 
 \nonumber \\
& & \qquad \times \left\langle
\left(\partial X^{\mu} (y_1) + i (2 \pi \alpha')^{\frac{1}{2}} p_{1 \nu} 
 \psi^{\nu} \psi^{\mu} (y_1) \right) S^{\dot{\alpha}} (y_2) 
S^{\dot{\beta}} (y_3) S_{\alpha} (z) S_{\beta} (\bar{z}) \prod_{j=1}^3 
e^{i \sqrt{2 \pi \alpha'} p_j \cdot X (y_j)} \right\rangle. \nonumber \\
\label{eq:amp1a}
\end{eqnarray}
We note that 
$\partial X^{\mu}$ in the last correlator of (\ref{eq:amp1a}) does not 
contribute to the amplitude because of symmetric property
of $\mathcal{F}^{(\alpha \beta) (A B)}$.
The correlation functions are calculated by using bosonization 
formulas summarized in Appendix A.
We then perform the world-sheet integral of the form
\begin{eqnarray}
\int^{\infty}_{- \infty} \! d y_2 \int^{y_2}_{- \infty} \! d y_3 \
\frac{(z - \bar{z})^2}{(y_2 - z) (y_2 - \bar{z}) (y_3 - z) (y_3 - 
\bar{z})} = (2 i)^2 \frac{\pi^2}{2}
\end{eqnarray}
which is done
 by fixing the world-sheet coordinates to $z = i, \bar{z} = - i, 
y_1 \to \infty$. 
The  amplitude becomes 
\begin{eqnarray}
& & \langle \! \langle V^{(0)}_A (p_1) V^{(-1/2)}_{\overline{\Lambda}} (p_2) 
 V^{(-1/2)}_{\overline{\Lambda}} (p_3) V^{(-1/2,-1/2)}_{\mathcal{F}} 
\rangle \! \rangle
 \nonumber \\
& & = - \frac{4 \pi^2 i}{k g^2_{\mathrm{YM}}} \mathrm{Tr} \left[
(\sigma^{\mu \nu})_{\alpha \beta} i p_{1 [\mu} A_{\nu]} (p_1) 
\overline{\Lambda}_{\dot{\alpha}A}
(p_2) \overline{\Lambda}^{\dot{\alpha}}_{\ B} (p_3) 
\right] (2 \pi \alpha')^{\frac{3}{2}} \mathcal{F}^{(\alpha \beta)(AB)}.
\label{eq:amp1aa}
\end{eqnarray}
The second term in (\ref{amp1}) can be evaluated in the same way.
The result is 
\begin{eqnarray}
& & \langle \! \langle V^{(0)}_H (p_1) V^{(-1/2)}_{\overline{\Lambda}} (p_2) 
 V^{(-1/2)}_{\overline{\Lambda}} (p_3) V^{(-1/2,-1/2)}_{\mathcal{F}} \rangle \! \rangle
 \nonumber \\
& & \qquad = - \frac{1}{2i} \frac{1}{2} \frac{8 \pi^2 i}{k g^2_{\mathrm{YM}}} \mathrm{Tr} \left[
(\sigma^{\mu \nu})_{\alpha \beta} H_{\mu \nu} (p_1) \overline{\Lambda}_{\dot{\alpha}A}
(p_2) \overline{\Lambda}^{\dot{\alpha}}_{\ B} (p_3) 
\right] (2 \pi \alpha')^{\frac{3}{2}} \mathcal{F}^{(\alpha \beta)(AB)}.
\label{eq:amp1b}
\end{eqnarray}
We need to add another color order contribution, which actually gives 
the same result and cancels the symmetric factor $1/2!$. 
The interaction terms  in the effective
Lagrangian
obtained from the amplitudes (\ref{eq:amp1aa}) and
(\ref{eq:amp1b}) are given by
\begin{eqnarray}
\mathcal{L}_1 &=&  \frac{4 \pi^2 i}{k g^2_{\mathrm{YM}}}
 \mathrm{Tr} \left[
(\sigma^{\mu \nu})_{\alpha \beta} \left(
\partial_{[\mu} A_{\nu]} - \frac{i}{2} H_{\mu \nu} 
\right) \overline{\Lambda}_{\dot{\alpha} A} 
\overline{\Lambda}^{\dot{\alpha}}_{\ B} 
\right] (2 \pi \alpha')^{\frac{3}{2}} \mathcal{F}^{(\alpha \beta) (AB)}.
\end{eqnarray}
\\
\underline{\textbullet $\langle \! \langle  V_{\Lambda} 
 V_{\overline{\Lambda}} V_{\varphi} V_{\mathcal{F}} \rangle \! \rangle$ 
 + $\langle \! \langle  V_{\Lambda} V_{\overline{\Lambda}} V_{H_{\varphi 
 \varphi}} V_{\mathcal{F}} \rangle \! \rangle$ }
\\
\\
The first term in the amplitudes (\ref{amp2}) is given by
\begin{eqnarray}
& & \langle \! \langle  V^{(-1/2)}_{\Lambda} (p_1) 
 V^{(-1/2)}_{\overline{\Lambda}} (p_2) V^{(0)}_{\varphi} (p_3)
V^{(-1/2,-1/2)}_{\mathcal{F}} \rangle \! \rangle
 \nonumber \\
& & = \frac{1}{2 \pi^2 \alpha'^2} \frac{1}{k g^2_{\mathrm{YM}}}
(2i) (2 \pi \alpha')^{3} \mathrm{Tr} \left[
\Lambda^{\gamma C} (p_1) \overline{\Lambda}_{\dot{\beta} 
D} (p_2) \varphi_a (p_3) \right] \mathcal{F}^{(\alpha \beta) (AB)} \nonumber \\
& & \qquad \times \int \! \frac{\prod_j d y_j d z d \bar{z}}{d V_{\mathrm{CKG}}} \
\langle e^{-\frac{1}{2} \phi (y_1)} e^{-\frac{1}{2} \phi (y_2)}
 e^{-\frac{1}{2} \phi (z)} e^{-\frac{1}{2} \phi (\bar{z})} \rangle 
 \nonumber \\
& & \qquad \frac{}{} \times \left\langle \frac{}{} S_{\gamma} S_C (y_1) S^{\dot{\alpha}} S^D (y_2)
\left(\partial X^a (y_3) + i (2 \pi \alpha')^{\frac{1}{2}} p_{3 \mu} 
 \psi^{\mu} \psi^a (y_3) \right) \right. \nonumber \\
& & \qquad \quad \left. \times S_{\alpha} S_A (z) S_{\beta} S_B (\bar{z}) 
	    \prod_{j=1}^3 e^{i \sqrt{2 \pi \alpha'} p_j \cdot X (y_j)} 
\right\rangle.
\end{eqnarray}
Here $\partial X^a$ does not contribute to the amplitude 
for the  (S,S)-type background.
Using the formula for the five point function of spin fields 
in Appendix A, we obtain
\begin{eqnarray}
& & \langle \! \langle V^{(-1/2)}_{\Lambda} (p_1) 
 V^{(-1/2)}_{\overline{\Lambda}} (p_2) V^{(0)}_{\varphi} (p_3)
V^{(-1/2,-1/2)}_{\mathcal{F}} \rangle \! \rangle
 \nonumber \\
& & =  \frac{4 \pi^2 i}{k g^2_{\mathrm{YM}}} 
\mathrm{Tr} \left[
\Lambda_{\alpha}^{\ C} (p_1) (\overline{\Sigma}^a)_{AC} 
\overline{\Lambda}_{\dot{\alpha} B} (p_2) (\sigma^{\mu})_{\beta}^{\ \dot{\alpha}}
i p_{3 \mu} \varphi_a (p_3) 
\right] (2 \pi \alpha')^{\frac{3}{2}} \mathcal{F}^{(\alpha \beta) (AB)}.
\end{eqnarray}
The amplitude which includes the auxiliary field $H_{\mu a}$ is given by
\begin{eqnarray}
& & \langle \! \langle  V^{(-1/2)}_{\Lambda} (p_1) 
 V^{(-1/2)}_{\overline{\Lambda}} (p_2) V^{(0)}_{H_{A \varphi}} (p_3)
V^{(-1/2,-1/2)}_{\mathcal{F}} \rangle \! \rangle 
 \nonumber \\
& & =  \frac{2}{2i} \frac{4 \pi^2 i}{k g^2_{\mathrm{YM}}} 
\mathrm{Tr} \left[  
\Lambda_{\alpha}^{\ C} (p_1) (\overline{\Sigma}^a)_{AC} 
\overline{\Lambda}_{\dot{\alpha} B} (p_2) (\sigma^{\mu})_{\beta}^{\ \dot{\alpha}}
H_{\mu a} (p_3)  
\right] (2 \pi \alpha')^{\frac{3}{2}} \mathcal{F}^{(\alpha \beta) (AB)}.
\end{eqnarray}
Another color order contribution needs to be added.
These amplitudes are obtained from the interaction 
\begin{eqnarray}
\mathcal{L}_2 &=&  \frac{4 \pi^2 i}{k g^2_{\mathrm{YM}}} 
\mathrm{Tr} \left[ \left\{  (\sigma^{\mu})_{\alpha}^{\ \dot{\alpha}} 
\left( \partial_{\mu} \varphi_a - i H_{\mu a} 
\right) , \overline{\Lambda}_{\dot{\alpha} A} \right\} 
(\overline{\Sigma}^a)_{BC}
 \Lambda_{\beta}^{\ C} \right] (2 \pi \alpha')^{\frac{3}{2}} 
\mathcal{F}^{(\alpha \beta) (AB)}.
\end{eqnarray}

\underline{\textbullet $\langle \! \langle \varphi \varphi \varphi 
\mathcal{F} \rangle \! \rangle +
\langle \! \langle H_{A \varphi} \varphi \varphi \mathcal{F} \rangle \! 
\rangle +
  \langle \! \langle H_{A \varphi} H_{A \varphi} \varphi 
\mathcal{F} \rangle \! \rangle $}
\\
\\
The first term in (\ref{amp3}) is given by
\begin{eqnarray}
& & \langle \! \langle V^{(0)}_{\varphi} (p_1) V^{(0)}_{\varphi} (p_2) 
V^{(-1)}_{\varphi} (p_3) V^{(-1/2,-1/2)}_{\mathcal{F}} 
\rangle \! \rangle
\nonumber \\
& & = \frac{1}{2 \pi^2 \alpha'^2} \frac{1}{k g^2_{\mathrm{YM}}}
(2i)^2 \frac{1}{\sqrt{2}} (2 \pi \alpha')^{\frac{5}{2}} \mathrm{Tr} \left[
\varphi_a (p_1) \varphi_b (p_2) \varphi_c (p_3) \right] 
\mathcal{F}^{(\alpha \beta) (AB)} \nonumber \\
& & \qquad \times \int \! \frac{\prod_j d y_j d z d \bar{z}}{d V_{\mathrm{CKG}}} \
\langle e^{- \phi (y_3)} e^{-\frac{1}{2} \phi (z)} e^{-\frac{1}{2} 
\phi (\bar{z})} \rangle 
 \nonumber \\
& & \qquad \times \frac{}{} \left\langle \left( \partial X^a (y_1) + i (2 \pi \alpha')^{\frac{1}{2}}
 p_{1 \mu} \psi^{\mu} \psi^a (y_1) \right) \left(
\partial X^b (y_2) + i (2 \pi \alpha')^{\frac{1}{2}} p_{2 \nu} \psi^{\nu}
\psi^{b} (y_2) \right) \right. \nonumber \\
& & \qquad \times \left. \psi^c (y_3) S_{\alpha} (z) S_{\beta} (\bar{z}) S_A (z) 
S_{\beta} (\bar{z}) S_B (\bar{z}) \prod_{j=1}^3 e^{i \sqrt{2 \pi \alpha'} 
p_j \cdot X (y_j)} \right\rangle.
\end{eqnarray}
In the above amplitudes
the term containing $\partial X^a \partial X^b$ 
gives the contribution 
 $\langle S_{\alpha} S_{\beta} 
\rangle \sim \varepsilon_{\alpha \beta}$, which becomes zero 
after the contraction with the (S,S)-type background. 
The terms containing the single $\partial X$ 
do not contribute to the amplitude
due to $\langle S_{\alpha} 
\psi^{\mu} S_{\beta} \rangle = 0$.
We obtain
\begin{eqnarray}
& & \langle \! \langle V^{(0)}_{\varphi} (p_1) V^{(0)}_{\varphi} (p_2) 
V^{(-1)}_{\varphi} (p_3) V^{(-1/2,-1/2)}_{\mathcal{F}} \rangle \! \rangle
\nonumber \\ 
& & = - \frac{4 \pi^2}{k g^2_{\mathrm{YM}}} \mathrm{Tr} 
\left[ (\sigma^{\mu \nu})_{\alpha \beta} 
 (\overline{\Sigma}^a \Sigma^b \overline{\Sigma}^c)_{AB} i p_{1 \mu} 
\varphi_a (p_1) i p_{2 \nu} \varphi_b (p_2) \varphi_c (p_3) 
\right] (2 \pi \alpha')^{\frac{3}{2}} \mathcal{F}^{(\alpha \beta)(AB)}.
\end{eqnarray}
The amplitudes including auxiliary fields can be calculated in a similar
way. Multiplying appropriate weight and symmetric factors, the interaction 
terms in the Lagrangian are shown to become
\begin{eqnarray}
\mathcal{L}_3 &=&  \frac{1}{3} \frac{4 \pi^2}{k g^2_{\mathrm{YM}}} 
\mathrm{Tr}
\left[ (\sigma^{\mu \nu})_{\alpha \beta}  (\overline{\Sigma}^a \Sigma^b 
\overline{\Sigma}^c)_{AB}
\partial_{\mu} \varphi_a  \partial_{\nu} \varphi_b \varphi_c 
\right] (2 \pi \alpha')^{\frac{3}{2}} \mathcal{F}^{(\alpha \beta) (AB)} 
\nonumber \\
& & + \frac{1}{3} \frac{2}{2i} \frac{4 \pi^2}{k g^2_{\mathrm{YM}}} \mathrm{Tr}
\left[ (\sigma^{\mu \nu})_{\alpha \beta}  (\overline{\Sigma}^a \Sigma^b 
\overline{\Sigma}^c)_{AB}
\{ H_{\mu a}, \partial_{\nu}  \varphi_b \} \varphi_c 
\right] (2 \pi \alpha')^{\frac{3}{2}} \mathcal{F}^{(\alpha \beta) (AB)} 
\nonumber \\
& & + \frac{1}{3} \frac{2^2}{(2i)^2} \frac{4 \pi^2}{k g^2_{\mathrm{YM}}}
 \mathrm{Tr}
\left[ (\sigma^{\mu \nu})_{\alpha \beta}  (\overline{\Sigma}^a \Sigma^b 
\overline{\Sigma}^c)_{AB}
H_{\mu a}  H_{\nu b} \varphi_c  
\right] (2 \pi \alpha')^{\frac{3}{2}} \mathcal{F}^{(\alpha \beta) (AB)}.
\end{eqnarray}
\\
\\
\underline{\textbullet $\langle \! \langle A_{\mu} H_{ab} \varphi 
\mathcal{F} \rangle \! \rangle + \langle \! \langle
H_{\mu \nu} H_{ab} \varphi \mathcal{F} \rangle \! \rangle $ }
\\
\\
Now we compute the amplitude (\ref{amp4}).
The first term in (\ref{amp4}) is given by
\begin{eqnarray}
& & \langle \! \langle V^{(0)}_{A} (p_1) V^{(0)}_{H_{\varphi \varphi}} (p_2) 
V^{(-1)}_{\varphi} (p_3) V^{(-1/2,-1/2)}_{\mathcal{F}} \rangle \! 
\rangle
\nonumber \\
& & = \frac{1}{2 \pi^2 \alpha'^2} \frac{1}{k g^2_{\mathrm{YM}}}
(2i) \left( -\frac{1}{\sqrt{2}} \right) \frac{1}{\sqrt{2}} (2 \pi
\alpha')^{3} 
\mathrm{Tr} \left[
A_{\mu} (p_1) H_{ab} (p_2) \varphi_c (p_3) \right] 
\mathcal{F}^{(\alpha \beta) (AB)} \nonumber \\
& & \qquad \times \int \! \frac{\prod_j d y_j d z d \bar{z}}{d V_{\mathrm{CKG}}} \
\langle e^{- \phi (y_3)} e^{-\frac{1}{2} \phi (z)} e^{-\frac{1}{2} 
\phi (\bar{z})} \rangle 
 \langle \psi^a \psi^b (y_2) \psi^c (y_3) S_A (z) S_B (\bar{z}) \rangle 
 \nonumber \\
& & \qquad \times \left\langle \left( \partial X^{\mu} (y_1) 
+ i (2 \pi \alpha')^{\frac{1}{2}}
p_{1 \nu} \psi^{\nu} \psi^{\mu} (y_1) 
\right) S_{\alpha} (z) S_{\beta} (\bar{z}) \prod_{j=1}^3 e^{i \sqrt{2 
\pi \alpha'} p_j \cdot X (y_j)} \right\rangle.
\end{eqnarray}
The term including $\partial X$  does not contribute to the amplitude
 for the (S,S)-type background again. 
Evaluating the correlation functions, we obtain
\begin{eqnarray}
& & \langle \! \langle V^{(0)}_{A} (p_1) V^{(0)}_{H_{\varphi \varphi}} (p_2) 
V^{(-1)}_{\varphi} (p_3) 
V^{(-1/2,-1/2)}_{\mathcal{F}} \rangle \! \rangle
\nonumber \\
& & =  \frac{i \sqrt{2} \pi^2}{k g^2_{\mathrm{YM}}} \mathrm{Tr} \left[
(\sigma^{\mu \nu})_{\alpha \beta} (\overline{\Sigma}^a \Sigma^b 
\overline{\Sigma}^c)_{AB} 
i p_{1 \mu} A_{\nu} (p_1) H_{ab} (p_2) 
\varphi_c (p_3) 
\right] (2 \pi \alpha')^{\frac{3}{2}} \mathcal{F}^{(\alpha \beta)(AB)}.
\nonumber \\
\end{eqnarray}
Taking into account other color ordered contributions and adding the 
second term 
in (\ref{amp4}), the interaction terms become
\begin{eqnarray}
\mathcal{L}_4 =  - \frac{\sqrt{2} \pi^2 i}{k g^2_{\mathrm{YM}}} \mathrm{Tr}
\left[ (\sigma^{\mu \nu})_{\alpha \beta} (\overline{\Sigma}^a 
\Sigma^b \overline{\Sigma}^c)_{AB}
\left(\partial_{[\mu} A_{\nu]} - \frac{i}{2} H_{\mu \nu} \right) 
\{ H_{ab}, \varphi_c \} 
\right] (2 \pi \alpha')^{\frac{3}{2}} \mathcal{F}^{(\alpha \beta) (AB)}.
\end{eqnarray}
\\
\underline{\textbullet $ \langle \! \langle H_{\varphi \varphi} \Lambda \Lambda 
\mathcal{F} \rangle \! \rangle$}
\\
\\
This amplitude is given by
\begin{eqnarray}
& & \langle \! \langle V^{(0)}_{\varphi \varphi} (p_1) 
V^{(-1/2)}_{\Lambda} (p_2) V^{(-1/2)}_{\Lambda} (p_3) 
V^{(-1/2,-1/2)}_{\mathcal{F}} \rangle \! \rangle \nonumber \\
& & = \frac{1}{2 \pi^2 \alpha'^2} \frac{1}{k g^2_{\mathrm{YM}}} (2 \pi 
\alpha')^{2 + \frac{3}{2}} \left( - \frac{1}{\sqrt{2}} \right) \mathrm{Tr} \left[ H_{ab} (p_1) 
\Lambda^{\gamma C} (p_2) \Lambda^{\delta D} (p_3) \right] 
\mathcal{F}^{(\alpha \beta) (AB)} \nonumber \\
& & \times \int \! \frac{\prod_j d y_j dz d \bar{z}}{d V_{\mathrm{CKG}}} 
\langle e^{- \frac{1}{2} \phi (y_2)} e^{- \frac{1}{2} \phi (y_3)} e^{- 
\frac{1}{2} \phi (z)} e^{- \frac{1}{2} \phi (\bar{z})} \rangle \nonumber \\
& & \times \langle \psi^a \psi^b (y_1) S_C (y_2) S_D (y_3) S_A (z) S_B 
(\bar{z}) \rangle \langle S_{\gamma} (y_2) S_{\delta} (y_3) S_{\alpha} 
(z) S_{\beta} (\bar{z}) \rangle.
\end{eqnarray}
Using the formula (\ref{eq:cor10}) in the appendix A, we get
\begin{eqnarray}
& & \langle \! \langle V^{(0)}_{\varphi \varphi} (p_1) 
V^{(-1/2)}_{\Lambda} (p_2) V^{(-1/2)}_{\Lambda} (p_3) 
V^{(-1/2,-1/2)}_{\mathcal{F}} \rangle \! \rangle \nonumber \\
& & \qquad \quad  = \frac{2 \sqrt{2} \pi^2 }{k g^2_{\mathrm{YM}}} \mathrm{Tr} \left[
(\overline{\Sigma}^{ab})_A^{\ A'} \varepsilon_{CD A'B} H_{ab} (p_1) 
\Lambda_{\alpha}^{\ C} (p_2) \Lambda_{\beta}^{\ D} (p_3) \right] (2 \pi \alpha')^{\frac{3}{2}}
\mathcal{F}^{(\alpha \beta) (AB)}.
\end{eqnarray}
Adding the color ordered amplitude and considering weight 
and phase factors, we find that the interaction term is
\begin{eqnarray}
\mathcal{L}_5 = - \frac{2 \sqrt{2} \pi^2 i }
{k g^2_{\mathrm{YM}}} \mathrm{Tr} \left[
(\overline{\Sigma}^{ab})_A^{\ A'} \varepsilon_{CD A'B} H_{ab}  
\Lambda_{\alpha}^{\ C}  
\Lambda_{\beta}^{\ D}  \right] (2 \pi \alpha')^{\frac{3}{2}} \mathcal{F}^{(\alpha \beta) (AB)}.
\end{eqnarray}

To summarize, the first order correction to the $\mathcal{N} 
= 4$ super Yang-Mills action
from the (S,S)-type graviphoton background is
\begin{eqnarray}
\mathcal{L}^{(1)}_{\mathrm{(S,S)}} = \mathcal{L}_1 +  
 \mathcal{L}_2 +  \mathcal{L}_3 + a_1 \mathcal{L}_4 + a_2 \mathcal{L}_5.
\end{eqnarray}
Here we have introduced additional weight factors $a_1$ and $a_2$,
which should be determined by higher point 
disk amplitudes without auxiliary fields.
In this paper we will determine these weight factors such that the
deformed Lagrangian is consistent with the one defined on $\mathcal{N}=1/2$
superspace.

By integrating out the auxiliary fields and defining the deformation 
parameter $C$ by
 $C^{(\alpha \beta) (AB)} \equiv - 8 \pi^2  (2 \pi \alpha')^{\frac{3}{2}} 
\mathcal{F}^{(\alpha \beta) (AB)}$, we find the deformed 
Lagrangian is expressed as
\begin{eqnarray}
\mathcal{L} = \mathcal{L}_{\mathrm{SYM}}^{\mathcal{N} = 4} + 
\mathcal{L}_{\mathrm{(S,S)}}^{(1)} + \mathcal{O} (C^2) \label{deformedN=4SYM}
\end{eqnarray}
where  $\mathcal{L}_{\mathrm{SYM}}^{\mathcal{N} = 4}$ is the ordinary 
$\mathcal{N} = 4$ super Yang-Mills action (\ref{N=4SYM}) and
\begin{eqnarray}
\mathcal{L}_{\textrm{(S,S)}}^{(1)} &=& - \frac{i}{2} 
\frac{1}{k g^2_{\mathrm{YM}}} 
\mathrm{Tr} \left[ F_{\mu \nu} \overline{\Lambda}_{\dot{\alpha} A} 
\overline{\Lambda}^{\dot{\alpha}}_{\ B}  \right] C^{\mu \nu (AB)} 
\nonumber\\
&&- \frac{i}{2} \frac{1}{k g^2_{\mathrm{YM}}} 
\mathrm{Tr} \left[ \left\{ D_{\mu} \varphi_a 
 , (\sigma^{\mu})_{\alpha \dot{\alpha}} 
\overline{\Lambda}^{\dot{\alpha}}_{\ A} \right\} (\overline{\Sigma}^a)_{BC}
 \Lambda_{\beta}^{\ C} \right] C^{(\alpha \beta) (AB)} \nonumber \\
& & - \frac{1}{6} \frac{}{k g^2_{\mathrm{YM}}} 
\mathrm{Tr} \left[ (\sigma^{\mu \nu})_{\alpha \beta}  
	     (\overline{\Sigma}^a \Sigma^b \overline{\Sigma}^c)_{AB} 
D_{\mu} \varphi_a D_{\nu} \varphi_b \varphi_c \right] C^{(\alpha 
\beta) (AB)} \nonumber \\
& & - \frac{i}{3} \frac{a_1}{k g^2_{\mathrm{YM}}} 
\mathrm{Tr} \left[ \frac{}{} (\sigma^{\mu \nu})_{\alpha \beta} 
(\overline{\Sigma}^a
 \Sigma^b \overline{\Sigma}^c)_{AB} F_{\mu \nu} \varphi_a \varphi_b 
\varphi_c  \right] C^{(\alpha \beta) (AB)} \nonumber \\
& & - \frac{i}{4} \frac{a_2}{k g^2_{\mathrm{YM}}} \mathrm{Tr} \left[
(\overline{\Sigma}^{ab})_A^{\ A'} \varepsilon_{A'BCD} \varphi_a 
\varphi_b \Lambda_{\alpha}^{\ C} \Lambda_{\beta}^{\ D} 
\right] C^{(\alpha \beta) (AB)}.
\label{eq:lag1}
\end{eqnarray}
Here $C^{\mu \nu (AB)}$ is defined by $C^{\mu \nu (AB)} = 
C^{(\alpha \beta) (AB)} \varepsilon_{\beta \gamma} 
(\sigma^{\mu \nu})_{\alpha}^{\ \gamma}  $.
We note that the bosonic terms in (\ref{eq:lag1}) gives the
ones obtained from the Chern-Simons term \cite{My} in the R-R potential,
which was also observed from the $\mathcal{N}=4$ super Yang-Mills theory on
non(anti)commutative $\mathcal{N}=1$ superspace \cite{Im2} (see also
appendix B).
The reduction to deformed $\mathcal{N}=1$ superspace is done by
restriction of the deformation parameter $C^{(\alpha\beta)(AB)}$ to
$C^{(\alpha\beta)(11)}$.
The deformed Lagrangian (\ref{eq:lag1}) agrees with the 
non(anti)commutative one in \cite{Im2} if we choose the weight factors 
to be $a_1 = - \frac{1}{2}, a_2 = - 4 i$.

\section{Vacuum structure of deformed $\mathcal{N} = 4$ theory}
In this section we study
the vacuum structure of deformed $\mathcal{N} = 4$ SYM 
theory  based on the Lagrangian (\ref{deformedN=4SYM}).
For simplicity, we take $\Lambda = \overline{\Lambda} = 0$, and consider 
$\varphi_a$ as constants.
 We also assume that only $U(1)$ part of 
gauge field strength $F^{U(1)}_{\mu \nu}$ is non-vanishing 
constant. 
In this case, 
the Lagrangian becomes
\begin{eqnarray}
\mathcal{L}_{\mathrm{scalar}} = \frac{1}{k g^2_{\mathrm{YM}}} \mathrm{Tr}
\left[ \frac{1}{4} [\varphi_a, \varphi_b]^2  + \frac{i}{6} 
(\overline{\Sigma}^a \Sigma^b 
\overline{\Sigma}^c )_{AB} F_{\mu \nu}^{U(1)} C^{\mu \nu (AB)} 
\varphi_a \varphi_b \varphi_c 
\right].
\end{eqnarray}
The equation of motion is given by
\begin{eqnarray}
[\varphi_b, [\varphi_a, \varphi_b] ] + \frac{i}{4} (\overline{\Sigma}^a 
\Sigma^b 
 \overline{\Sigma}^c)_{AB} F_{\mu \nu}^{U(1)} C^{\mu \nu (AB)} 
 [\varphi_b, \varphi_c] = 0. \label{eqofm_scalar}
\end{eqnarray}
We want to find the solution with the ansatz
\begin{eqnarray}
& & [\varphi_{\hat{a}}, \varphi_{\hat{b}}] = i \kappa 
 \varepsilon_{\hat{a} \hat{b} \hat{c}} \varphi_{\hat{c}}, \quad 
 (\hat{a}, \hat{b}, \hat{c} =1,2,3), \nonumber \\
& & \varphi_{\hat{i}} = 0 \quad (\hat{i} = 4,5,6). \label{ansatz}
\end{eqnarray}
Taking the contraction with $C^{(AB)}$, totally antisymmetric part
of $(\overline{\Sigma}^a \Sigma^b \overline{\Sigma}^c)_{AB}$ remains.
We find that the equation (\ref{eqofm_scalar}) reduces to
\begin{eqnarray}
[\varphi_{\hat{b}}, [\varphi_{\hat{a}}, \varphi_{\hat{b}}]] 
+ \frac{i}{4} \varepsilon_{\hat{a}\hat{b}\hat{c}} (F \cdot C) 
 [\varphi_{\hat{b}}, \varphi_{\hat{c}}] = 0 \label{eom_fuzzy}
\end{eqnarray}
where 
$(\overline{\Sigma}^{\hat{a}} \Sigma^{\hat{b}} 
\overline{\Sigma}^{\hat{c}})_{AB} C^{\mu \nu (AB)}$ is
written as 
$ \varepsilon^{\hat{a} \hat{b} \hat{c}} 
M_{AB} C^{\mu \nu (AB)}$ 
for a symmetric matrix $M_{AB}$ 
and $F \cdot C \equiv F^{U(1)}_{\mu \nu} 
C^{\mu \nu (AB)} M_{AB}$. 
Applying the ansatz (\ref{ansatz}) we find 
(\ref{eqofm_scalar}) can be rewritten as 
\begin{eqnarray}
\left( \kappa^2 + \frac{1}{4} (F \cdot C) \kappa  \right) 
 \varepsilon_{\hat{a} \hat{b} \hat{c}} 
 \varepsilon_{\hat{b} \hat{c} \hat{d}} \varphi_{\hat{d}} = 0.
\end{eqnarray}
Thus the constant $\kappa$ should be
\begin{eqnarray}
\textrm{(i)} & & \kappa = 0, \\
\textrm{(ii)} & & \kappa = - \frac{1}{4} (F \cdot C).
\end{eqnarray}
In the case (i) this gives the ordinary commutative configuration of 
D3-branes. 
However, in  the case (ii), 
due to the non-zero R-R background $C$, we have the fuzzy two-sphere 
configuration $[\varphi_{\hat{a}}, \varphi_{\hat{b}}] = i \kappa 
\varepsilon_{\hat{a} \hat{b} \hat{c}} \varphi_{\hat{c}}$.
Here we regard
$\varphi_{\hat{a}}$ as the generators of $SU(2)$ subalgebra embedded 
in the $N$-dimensional matrix representation of the gauge group
$U(N)$, which are normalized as $\varphi^{\hat{a}} \varphi^{\hat{a}} 
= t \mathbf{1}_{N \times N}$. 
The radius of the fuzzy two-sphere is given by
\begin{eqnarray}
R^2 \equiv  \varphi_{\hat{a}}^2 = \kappa^2 t \mathbf{1}_{N \times N}.
\end{eqnarray} 

\section{Conclusions and discussion}
In this paper, we have investigated the effects of constant self-dual
R-R graviphoton 5-form 
background to the $U(N)$ $\mathcal{N} = 4$ super Yang-Mills
 theory defined on D3-brane 
world-volume.
In this paper we discussed the first order correction from the
graviphoton background 
to the $\mathcal{N} = 4$ super Yang-Mills theory keeping 
$(2 \pi \alpha')^{\frac{3}{2}} \mathcal{F} = C$ 
fixed in the zero-slope limit. 
This scaling gives the same dimension with the non(anti)commutativity 
parameter of superspace.
The deformed action would be defined on the 
non(anti)commutative $\mathcal{N} = 4$ 
superspace which is characterized by the Clifford 
algebra $\{ \theta^{\alpha A}, \theta^{\beta B} \} = C^{(\alpha \beta) 
(AB)}$. 
It would be interesting to study this deformation by using pure-spinor
formalism \cite{Be} of superstring, which provides a useful method to 
studying higher order corrections.

By restricting the R-R 5-form field strength to the $\mathcal{N} = 1$ 
deformation parameter and assigning appropriate weight factors 
to the amplitudes,
we found that the effective action agrees with the 
one defined on $\mathcal{N} = 1/2$
superspace \cite{Im2} in our convention.
We also find the fuzzy two-sphere vacuum configuration which is induced 
by non-zero R-R background as in \cite{My}.
In our calculation, there are no tadpole contribution nor divergent 
structure of the disk amplitudes which suggest the consistency of the 
constant R-R 5-form background with flat space-time. 

We can do similar calculations for the other 
types of R-R background such as (S,A), (A,S), (A,A).
As pointed out in \cite{ItSa}, the (S,A) and (A,S)-type backgrounds can 
not be interpreted as ordinary 
non(anti)commutative deformation of superspace 
because their index structures are different. On the other hand, (A,A)-type 
background, which corresponds to R-R 1-form background, would 
provide the non(anti)commutative deformation of superspace.

Another interesting issue is 
to choose different scaling conditions for the R-R background 
in the zero-slope limit.
For example, in $\mathcal{N} = 2$ case, the (S,A)-type background 
with the scaling $(2 \pi \alpha')^{\frac{1}{2}} \mathcal{F} = C$ is 
studied in \cite{Billo-N=2}. 
The R-R three-form is regarded as the $\Omega$-background,
which was used for the integration over the instanton moduli space
\cite{LoMaNe}.
The C-deformation scaling $(2 \pi \alpha')^{- \frac{1}{2}} \mathcal{F} = 
C$ would be also interesting \cite{OoVa}.

\subsection*{Acknowledgments}
We would like to thank A.~Lerda for useful comments.
The work of K.~I. is  supported in part by the Grant-in-Aid for Scientific 
Research No. 18540255 from Ministry of Education, Science, 
Culture and Sports of Japan. S.~S. is supported by Yukawa Memorial
Foundation.

\begin{appendix}

\section{$\mathcal{N} = 4$ effective rules}
In this appendix we summarize some definitions and useful formulas which 
appear in this paper.

We define spin fields in six dimensions by
\begin{eqnarray}
& & S^1 = e^{\frac{1}{2} \phi_3 + \frac{1}{2} \phi_4 + \frac{1}{2} 
 \phi_5}, \qquad S_1 = e^{- \frac{1}{2} \phi_3 - \frac{1}{2} \phi_4 - 
 \frac{1}{2} \phi_5}, \nonumber \\
& & S^2 = i e^{\frac{1}{2} \phi_3 - \frac{1}{2} \phi_4 - \frac{1}{2} 
 \phi_5}, \qquad S_2 = i e^{- \frac{1}{2} \phi_3 + \frac{1}{2} \phi_4 + 
 \frac{1}{2} \phi_5}, \nonumber \\
& & S^3 = i^2 e^{- \frac{1}{2} \phi_3 + \frac{1}{2} \phi_4 - \frac{1}{2} 
v \phi_5}, \qquad S_3 = i^2 e^{\frac{1}{2} \phi_3 - \frac{1}{2} \phi_4 + 
 \frac{1}{2} \phi_5}, \nonumber \\
& & S^4 = i^3 e^{- \frac{1}{2} \phi_3 - \frac{1}{2} \phi_4 + \frac{1}{2} 
 \phi_5}, \qquad S_4 = i^3 e^{\frac{1}{2} \phi_3 
+ \frac{1}{2} \phi_4 - \frac{1}{2} \phi_5}.
\end{eqnarray}

The correlation functions for ten-dimensional spin fields can be
realized 
as the product of four-dimensional correlator and six-dimensional
ones.
Each correlator is expressed in terms of gamma matrices, which is 
evaluated by using the effective rules listed below.

We firstly write down the correlation functions for four-dimensional 
spin fields:
\begin{eqnarray}
\langle S_{\alpha} (z) S_{\beta} (\bar{z}) \rangle = \varepsilon_{\alpha 
\beta} (z - \bar{z})^{- \frac{1}{2}},
\end{eqnarray}
\begin{eqnarray}
\langle S^{\dot{\alpha}} (y_1) S^{\dot{\beta}} (y_2) \rangle = 
\varepsilon^{\dot{\alpha} \dot{\beta}} (y_1 - y_2)^{-\frac{1}{2}},
\end{eqnarray}
\begin{eqnarray}
\langle  S^{\dot{\alpha}} (y_1) S^{\dot{\beta}} (y_2) 
S_{\alpha} (z) S_{\beta} (\bar{z}) \rangle = \varepsilon^{\dot{\alpha} 
\dot{\beta}} \varepsilon_{\alpha 
\beta} (y_1 - y_2)^{-\frac{1}{2}} (z - \bar{z})^{- \frac{1}{2}},
\end{eqnarray}
\begin{eqnarray}
\langle S^{\alpha} (y_1) S^{\beta} (y_2) S^{\gamma} (z) S^{\delta} 
(\bar{z}) \rangle 
&=& \left[ (y_1 - y_2) (y_1 - z) (y_1 - \bar{z}) (y_2 - z) (y_2 - \bar{z}) 
 (z - \bar{z}) \right]^{-\frac{1}{2}} \nonumber \\
& & \qquad \times \left[ \varepsilon^{\alpha \delta} 
\varepsilon^{\beta \gamma} 
(y_1 - z) (y_2 - \bar{z}) - \varepsilon^{\alpha \gamma} 
\varepsilon^{\beta \delta} (y_2 - z) (y_1 
- \bar{z}) \right] \nonumber \\
&=& \left[ (y_1 - y_2) (y_1 - z) (y_1 - \bar{z}) (y_2 - z) (y_2 - \bar{z}) 
 (z - \bar{z}) \right]^{-\frac{1}{2}} \nonumber \\
& & \qquad \times \left[ - \varepsilon^{\alpha \beta} 
 \varepsilon^{\gamma \delta} 
(y_1 - \bar{z}) (y_2 - z) + \varepsilon^{\alpha \delta} 
\varepsilon^{\beta \gamma} (y_1 - y_2) 
(z - \bar{z}) \right]. \nonumber \\
\end{eqnarray}
The correlators including world-sheet fermions become for example 
\begin{eqnarray}
\langle S^{\dot{\alpha}} (y_1) \psi^{\mu} (y_2) S_{\alpha} (y_3) \rangle
= \frac{1}{\sqrt{2}} (\bar{\sigma}^{\mu})^{\dot{\alpha}}_{\ \alpha} 
(y_1 - y_2)^{- \frac{1}{2}} (y_2 - y_3)^{-\frac{1}{2}},
\end{eqnarray}
\begin{eqnarray}
& & \langle \psi^{\mu} \psi^{\nu} (y_1) S^{\dot{\alpha}} (y_2) S^{\dot{\beta}} 
 (y_3) S_{\alpha} (z) S_{\beta} (\bar{z}) \rangle \nonumber \\
& & = (y_2 - y_3)^{- \frac{1}{2}} (z - \bar{z})^{- \frac{1}{2}}
\left[ (\bar{\sigma}^{\mu \nu})^{\dot{\alpha} \dot{\beta}} 
 \varepsilon_{\alpha \beta} \frac{(y_2 - y_3)}{(y_1 - y_2) (y_1 - y_3)}
+ (\sigma^{\mu \nu})_{\alpha \beta} \varepsilon^{\dot{\alpha} \dot{\beta}}
\frac{(z - \bar{z})}{(y_1 - z) (y_1 - \bar{z})} \right]. \nonumber \\
\end{eqnarray}
Next, the correlators for six-dimensional spin fields used in this
paper are
\begin{eqnarray}
& & \langle S^A (z) S_B (w) \rangle =  \delta^A_{\ B} (z - 
 w)^{-\frac{3}{4}}, \nonumber \\
& & \langle S^A (z) S^B (w) \rangle = \langle S_A (z) S_B (w) \rangle = 0,
\end{eqnarray}
\begin{equation}
\langle S_{A} (z_1) S_{B} (z_2) S_{C} (z_3) S_{D} (z_4) \rangle
= \frac{ \epsilon_{ABCD}}
{
(z_{1} - z_{2})^{\frac{1}{4}}(z_{1} - z_{3})^{\frac{1}{4}}
(z_{1} - z_{4})^{\frac{1}{4}}(z_{2} - z_{3})^{\frac{1}{4}}
(z_{2} - z_{4})^{\frac{1}{4}}(z_{3} - z_{4})^{\frac{1}{4}}
}
\end{equation}
and 
\begin{eqnarray}
& & \langle S^A (y_1) S^B (y_2) S_C (y_3) S_D (y_4) \rangle  \nonumber \\
& & \qquad = (z_1 - z_2)^{- \frac{1}{4}} (z_1 - z_3)^{- \frac{3}{4}} 
(z_1 - z_4)^{- \frac{3}{4}} (z_2 - z_3)^{- \frac{3}{4}} 
(z_2 - z_4)^{- \frac{3}{4}}
(z_3 - z_4)^{-\frac{1}{4}}
\nonumber \\
& & \qquad \qquad \times \left[
- (z_1 - z_4) (z_2 - z_3)  \delta^A_{\ C} \delta^B_{\ D} 
+ (z_1 - z_3) (z_2 - z_4) \delta^A_{\ D} \delta^B_{\ C} 
\right].
\end{eqnarray}
Here $\epsilon_{ABCD}$ is an anti-symmetric tensor with $\epsilon_{1234} = 1$.
The correlators including world-sheet fermions are
\begin{eqnarray}
\langle \psi^a (y_1) S_A (z) S_B (\bar{z}) \rangle = \frac{1}{\sqrt{2}} 
\left( \overline{\Sigma}^a \right)_{AB} (y_1 - z)^{- \frac{1}{2}} (y_1 - 
\bar{z})^{-\frac{1}{2}} (z - \bar{z})^{- \frac{1}{4}},
\end{eqnarray}
\begin{eqnarray}
& & \langle \psi^a \psi^b (y_1) \psi^c (y_2) S_A (z) S_B (\bar{z}) \rangle
\nonumber \\
& & = + \frac{1}{\sqrt{2}} \frac{1}{y_1 - y_2} \left[ \delta^{ac} 
(\overline{\Sigma}^b)_{AB}
(y_2 - z)^{-\frac{1}{2}} (y_2 - \bar{z})^{-\frac{1}{2}} (z - 
\bar{z})^{\frac{1}{4}} \right] \nonumber \\
& & \quad + \frac{1}{\sqrt{2}} \frac{1}{y_1 - y_2} \left[ \delta^{bc} 
(\overline{\Sigma}^a)_{AB}
(y_2 - z)^{-\frac{1}{2}} (y_2 - \bar{z})^{-\frac{1}{2}} (z - 
\bar{z})^{\frac{1}{4}} \right] \nonumber \\
& & \quad - \frac{1}{2\sqrt{2}} \frac{1}{y_1 - z} 
\left[ (\overline{\Sigma}^{ab})_A^{\ B'}
(\overline{\Sigma}^c)_{B'B} (y_2 - z)^{-\frac{1}{2}} 
(y_2 - \bar{z})^{-\frac{1}{2}} (z - 
\bar{z})^{\frac{1}{4}} \right] \nonumber \\
& & \quad + \frac{1}{2\sqrt{2}} \frac{1}{y_1 - \bar{z}} \left[ 
(\overline{\Sigma}^{ab})_B^{\ B'}
(\overline{\Sigma}^c)_{AB'} (y_2 - z)^{-\frac{1}{2}} 
(y_2 - \bar{z})^{-\frac{1}{2}} (z - 
\bar{z})^{\frac{1}{4}} \right],
\end{eqnarray}
where we have defined $(\overline{\Sigma}^{ab})_{A}^{\ B} = \frac{1}{4} 
\left( (\overline{\Sigma}^a)_{AC} (\Sigma^b)^{CB} - (\overline{\Sigma}^b)_{AC}
 (\Sigma^a)^{CB}  \right)$. 
The following formulas are valid only 
 when they are contracted with the (S,S)-type background $C^{(\alpha 
 \beta) (AB)}$:
\begin{eqnarray}
& & \langle S_{\gamma} S_C (y_1) S^{\dot{\alpha}} S^D (y_2)
\psi^{\mu} \psi^a (y_3) S_{\alpha} S_A (z) S_{\beta} S_D (\bar{z})
\rangle \nonumber \\
& & =  \frac{1}{2} \varepsilon_{\gamma \beta} 
 (\sigma^{\mu})_{\alpha}^{\ \dot{\alpha}} (\overline{\Sigma}^a)_{AC} 
 \delta^D_{\ B} (y_1 - y_2)^{- \frac{3}{4}} (y_1 - z)^{- \frac{3}{4}}
(y_1 - \bar{z})^{- \frac{3}{4}} (y_1 - y_2) \nonumber \\
& & \qquad \times (y_2 - z)^{- \frac{3}{4}} (y_2 - \bar{z})^{- \frac{3}{4}}
(y_3 - z)^{-1} (y_3 - \bar{z})^{-1} (z - \bar{z})^{- \frac{3}{4}} 
(z - \bar{z})^2,
\end{eqnarray}
\begin{eqnarray}
& & \langle \psi^{\mu} \psi^a (y_1) \psi^{\nu} \psi^b (y_2) \psi^c (y_3)
S_{\alpha} S_A (z) S_{\beta} S_B (\bar{z}) \rangle \nonumber \\
& & = - \frac{1}{4 \sqrt{2}} (\sigma^{\mu})_{\alpha \dot{\alpha}} 
 (\bar{\sigma}^{\nu})^{\dot{\alpha}}_{\ \beta} (\overline{\Sigma}^a 
 \Sigma^b \overline{\Sigma}^c)_{AB} \nonumber \\
& & \qquad \times (y_1 - z)^{-1} (y_1 - \bar{z})^{-1}
(y_2 - z)^{-1} (y_2 - \bar{z})^{-1}
(y_3 - z)^{-\frac{1}{2}} (y_3 - 
\bar{z})^{- \frac{1}{2}} (z - \bar{z})^{\frac{5}{4}}.
\end{eqnarray}
\begin{eqnarray}
& & \langle \psi^a \psi^b (y_1) S_C (y_2) S_D (y_3) S_A (z) S_B (\bar{z}) 
\rangle \nonumber \\
& & = (\overline{\Sigma}^{ab})_A^{\ A'} \varepsilon_{CDA'B} 
\frac{z - \bar{z}}{(y_1 - z) (y_1 - \bar{z})}
\left[
(y_2 - y_3) (y_2 - z) (y_2 - \bar{z}) (y_3 - z) (y_3 - \bar{z}) (z - 
\bar{z}) \right]^{-\frac{1}{4}}. \nonumber \\
\label{eq:cor10}
\end{eqnarray}

\section{$\mathcal{N} = 4$ super Yang-Mills theory on 
$\mathcal{N} = 1/2$  superspace}
In this appendix we calculate the Lagrangian of 
$\mathcal{N} = 4$ super Yang-Mills theory
defined on $\mathcal{N} = 1/2$  superspace.
In terms of $\mathcal{N}=1$ superfields, this theory is constructed by
a vector superfield $V(x,\theta,\bar{\theta})$ and three 
chiral superfields $\Phi_i(y,\theta)$ $(i=1,2,3)$ which belong to the 
adjoint representation of the gauge group $U(N)$.
The deformed Lagrangian \cite{Im2} is defined by 
\begin{eqnarray}
\mathcal{L}^{\mathcal{N} = 4}_c &=& \frac{1}{k} \int \! d^2 \theta d^2 
\bar{\theta} \ 
 \mathrm{Tr} \sum^3_{i=1} \left( \overline{\Phi}_i * e^V * \Phi_i * 
 e^{-V}  \right) \nonumber \\
& & \qquad + \frac{1}{16 k g^2_{\mathrm{YM}}} \int \! d^2 \theta \ \mathrm{Tr} 
\left( W^{\alpha} * W_{\alpha} \right) + \frac{1}{16 k
g^2_{\mathrm{YM}}} 
\int \! d \bar{\theta} \ 
\mathrm{tr} \left( \overline{W}_{\dot{\alpha}} * 
\overline{W}^{\dot{\alpha}} \right)
\nonumber \\
& & \qquad - \frac{\sqrt{2}}{3} \frac{g_{\mathrm{YM}}}{k} \int \! d^2 \theta \ 
\mathrm{Tr} \varepsilon^{ijk} \left( \Phi_i * \Phi_j * \Phi_k \right)
+ \frac{\sqrt{2}}{3} \frac{g_{\mathrm{YM}}}{k} \int \! d^2 \bar{\theta} \
\mathrm{Tr} \varepsilon^{ijk} \left( \overline{\Phi}_i * \overline{\Phi}_j 
	     * \overline{\Phi}_k \right).
\nonumber \\
\label{eq:lag10}
\end{eqnarray}
Here the star product is defined by $ f(\theta) * g (\theta) = f 
(\theta) \exp\left[ - \frac{1}{2} C^{\alpha \beta} 
\overleftarrow{Q_\alpha} 
\overrightarrow{Q_\beta} \right] g 
(\theta)$.
$Q_\alpha$ is the supercharge defined on the superspace. 
It is convenient to redefine
the component fields of a superfield such that they transform
canonically under the gauge transformation.
The expansion of the chiral superfield is the same as the undeformed one:
\begin{eqnarray}
\Phi_i (y, \theta) = \phi_i (y) + i \sqrt{2} \theta \psi_i (y) +  
\theta \theta F_i (y).
\end{eqnarray}
The anti-chiral superfield 
is expanded as \cite{ArItOh}
\begin{eqnarray}
\overline{\Phi}_i (\bar{y}, \bar{\theta}) = 
\bar{\phi}_i (\bar{y}) + i \sqrt{2} \bar{\theta} \bar{\psi}_i (\bar{y})
+ \bar{\theta} \bar{\theta} \left( \bar{F}_i (\bar{y}) + i C^{\mu \nu} 
\partial_{\mu} \{ \bar{\phi}_i, A_{\nu} \} (\bar{y})
 - \frac{g_{\mathrm{YM}}}{2}
C^{\mu \nu} [A_{\mu}, \{ A_{\nu}, \bar{\phi}_i \}] (\bar{y})  \right)
\nonumber \\
\end{eqnarray}
where we have defined $C^{\mu \nu} = C^{\alpha \beta} 
\varepsilon_{\beta \gamma} (\sigma^{\mu \nu})_{\alpha}^{\ \gamma} $.
The vector superfield in the Wess-Zumino gauge is \cite{Se}
\begin{eqnarray}
V (y, \theta, \bar{\theta}) &=& - \theta \sigma^{\mu} \bar{\theta}
A_{\mu} (y) + i \theta \theta \bar{\theta} \bar{\lambda} (y) 
- i \bar{\theta} \bar{\theta} \theta^{\alpha}
\left( \lambda_{\alpha} (y) + \frac{1}{4} \varepsilon_{\alpha \beta} 
C^{\beta \gamma} \sigma^{\mu}_{\gamma \dot{\gamma}} \{ 
\bar{\lambda}^{\dot{\gamma}}, A_{\mu} \} (y) \right) \nonumber \\
& & + \frac{1}{2} \theta \theta \bar{\theta} \bar{\theta} 
\left( D (y) - i \partial_{\mu} A^{\mu} (y) \right).
\end{eqnarray}
Rescaling appropriately component fields and 
 $C^{\alpha \beta}$ by gauge coupling 
constant $g_{\mathrm{YM}}$, we find that Lagrangian (\ref{eq:lag10})
becomes
\begin{eqnarray}
\mathcal{L}^{\mathcal{N} = 4}_c &=& \frac{1}{k g^2_{\mathrm{YM}}} 
\mathrm{Tr} \left[ - \frac{1}{4} F_{\mu \nu} F^{\mu \nu} - \frac{1}{4} 
\tilde{F}^{\mu \nu} F_{\mu \nu} - D^{\mu} \bar{\phi}_i D_{\mu} \phi_i
+ \bar{F}_i F_i + \frac{1}{2} D^2 \right.
\nonumber \\
& & \qquad  - i \bar{\psi}_i \bar{\sigma}^{\mu} D_{\mu} \psi_i 
- i \bar{\lambda} \bar{\sigma}^{\mu} D_{\mu} \lambda 
 - i \sqrt{2} [\bar{\phi}_i, \psi_i] \lambda - i \sqrt{2} 
[\phi_i, \bar{\psi}_i] \bar{\lambda} + D [\phi_i, \bar{\phi}_i] 
\nonumber \\
& & \qquad - \frac{i}{2} C^{\mu \nu} F_{\mu \nu} \bar{\lambda} \bar{\lambda}
+ \frac{1}{8} |C|^2 (\bar{\lambda} \bar{\lambda})^2 + \frac{i}{2} C^{\mu 
\nu} F_{\mu \nu} \{ \bar{\phi}_i, F_i \} \nonumber \\
& & \qquad - \frac{\sqrt{2}}{2} C^{\alpha \beta} \{ D_{\mu} \bar{\phi}_i,
(\sigma^{\mu} \bar{\lambda} )_{\alpha} \} \psi_{i \beta} - \frac{1}{16} 
|C|^2 [\bar{\phi}_i, \lambda] [\bar{\lambda}, F_i] \nonumber \\
& & \qquad - \sqrt{2} \varepsilon^{ijk} 
\left( F_i \phi_j \phi_k - \phi_i \psi_j \psi_k
- \frac{1}{12} |C|^2 F_i F_j F_k - \frac{1}{2} C^{\alpha \beta} F_i 
\psi_{j \alpha} \psi_{k \beta} \right) \nonumber \\
& & \left. \qquad + \sqrt{2} \varepsilon^{ijk} \left( \bar{F}_i \bar{\phi}_j 
\bar{\phi}_k
- \bar{\phi}_i \bar{\psi}_j \bar{\psi}_k + \frac{2i}{3} C^{\mu \nu} 
F_{\mu \nu} \bar{\phi}_i \bar{\phi}_j \bar{\phi}_k + \frac{1}{3} C^{\mu \nu}
D_{\mu} \bar{\phi}_i D_{\nu} \bar{\phi}_j \bar{\phi}_k \right)
\right].
\nonumber \\
\end{eqnarray}
Integrating out the auxiliary fields we get
\begin{eqnarray}
\mathcal{L}^{\mathcal{N} = 4}_c &=& \frac{1}{k g^2_{\mathrm{YM}}} 
\mathrm{Tr} \left[ - \frac{1}{4} F_{\mu \nu} F^{\mu \nu} - \frac{1}{4} 
\tilde{F}^{\mu \nu} F_{\mu \nu} 
- i \bar{\lambda} \bar{\sigma}^{\mu} D_{\mu} \lambda 
 - i \bar{\psi}_i \bar{\sigma}^{\mu} D_{\mu} \psi_i 
- D^{\mu} \bar{\phi}_i D_{\mu} \phi_i  \right.  \nonumber \\
& & \qquad - i \sqrt{2} [\bar{\phi}_i, \psi_i] \lambda - i \sqrt{2} 
[\phi_i, \bar{\psi}_i] \bar{\lambda} - \frac{1}{2} [\phi_i, \bar{\phi}_i]^2
+ [\bar{\phi}_i, \bar{\phi}_j] [\phi_i, \phi_j] 
\nonumber \\
& & \qquad - \frac{i}{2} C^{\mu \nu} F_{\mu \nu} \bar{\lambda} \bar{\lambda}
+ \frac{1}{8} |C|^2 (\bar{\lambda} \bar{\lambda})^2 
- \frac{\sqrt{2}}{2} C^{\alpha \beta} \{ D_{\mu} \bar{\phi}_i,
(\sigma^{\mu} \bar{\lambda} )_{\alpha} \} \psi_{i \beta}  \nonumber \\
& &  \qquad - \sqrt{2} \varepsilon^{ijk} \left( - \phi_i \psi_j \psi_k 
+ \bar{\phi}_i \bar{\psi}_j \bar{\psi}_k + \frac{i}{3} C^{\mu \nu} 
F_{\mu \nu} \bar{\phi}_i \bar{\phi}_j \bar{\phi}_k - \frac{1}{3} C^{\mu \nu}
D_{\mu} \bar{\phi}_i D_{\nu} \bar{\phi}_j \bar{\phi}_k \right) \nonumber \\
& & 
\qquad - C^{\alpha \beta} [\bar{\phi}_i, \bar{\phi}_j] 
\psi_{i \alpha} 
\psi_{j \beta} 
+ \frac{\sqrt{2}}{16} |C|^2 \varepsilon^{ijk} [\bar{\phi}_i, \lambda] 
[\bar{\lambda}, \bar{\phi}_j \bar{\phi}_k]
\nonumber\\
& &\left. \qquad + \frac{1}{12} |C|^2 \varepsilon^{ipq} \varepsilon^{jrs}
[\bar{\phi}_i, \bar{\phi}_j] [\bar{\phi}_p, \bar{\phi}_q] [\bar{\phi}_r, 
\bar{\phi}_s] \right]. 
\end{eqnarray}
We note that the term 
$- C^{\alpha \beta} [\bar{\phi}_i, \bar{\phi}_j] \psi_{i \alpha} 
\psi_{j \beta}$ in the 5th line is absent in \cite{Im2}. 
The relation between scalar fields  $\varphi_a$ and 
$\phi_i$ is given by
\begin{eqnarray}
\varphi_{2 i - 1} = \frac{1}{\sqrt{2}} \left( \phi_i + \bar{\phi}_i \right),
\quad \varphi_{2i} = \frac{i}{\sqrt{2}} 
\left( \phi_i - \bar{\phi}_i \right), \quad (i=1,2,3).
\end{eqnarray}

\end{appendix}

\end{document}